\documentclass[preprint,pra]{revtex4-1}
\usepackage{graphicx}
\usepackage{dcolumn}
\usepackage{bm}
\usepackage{amsmath}
\usepackage{amssymb}
\usepackage{mathrsfs}
\usepackage{natbib}
\usepackage{subfig}

\begin{document}
\title{Controlling spontaneous emission of a two-level atom by hyperbolic metamaterials}

\author{Zheng Liu }

\author{Wei Li }
\email{waylee@mail.sim.ac.cn}
\author{Xunya Jiang }

\affiliation{State Key Laboratory of Functional Materials for Informatics, Shanghai
Institute of Microsystem and Information Technology, Chinese Academy of Sciences,
Shanghai 200050, China }

\begin{abstract}
Within the frame of  quantum optics  we  analyze  the properties of spontaneous emission
of two-level atom in media  with indefinite permittivity tensor where the geometry of the
dispersion relation is characterized by an ellipsoid or a hyperboloid(hyperbolic medium).
The decay rate is explicitly given with the orientation of the dipole transition matrix
element taken into account. It indicates that for the ellipsoid case the intensity of the
photons coupled into different modes can be tuned by changing the direction of the matrix
element and for the hyperboloid case it is found that spontaneous emission in hyperbolic
medium can be dramatically enhanced compared to the dielectric background. Moreover,
spontaneous emission exhibit the strong directivity and get the maximum in the asymptote
direction.
\end{abstract}

\maketitle

\section{introduction}
Metamaterials is composed of  the periodic dielectric element arrays where metal
mico-structure  e.g.the split-ring resonators(SRRs) is included in each cell  and used as
a effective continuous medium in a narrow microwave frequency
region\cite{contin1,contin2,contin3,contin4} or in the near-visible light
region\cite{contin5}.  The  materials have the  left-handed  property for its negative
permittivity and permeability simultaneously and  cause to the refocusing and phase
compensation\cite{veselago}, which  provide a new manipulating  space for the design of
quantum optical devices. For instance, the suppression of spontaneous emission and
superradiance over macroscopic distances in the left-handed materials
(LHM)\cite{superradiance}, the quantum interference enhancement between two spontaneous
emission transitions with the LHM\cite{interference}, long-lived entanglement between two
distant atoms via the LHM\cite{entanglement}. On the other hand the effective
permittivity or permeability of the metamaterials only holds for the specific
lpolarization and it mean that metamaterials is anisotropic as well as
left-handed\cite{indefinite,retrieval,chiral}. The anisotropy of metamaterials has also
been found some new optical properties  for example, the incident electric field
\textbf{E} can couple to the magnetic resonance of the SRRs when the electromagnetic
waves propagate perpendicular to the SRR plane\cite{coupling} and it can be utilized to
excite electron-spin resonance\cite{inverse}.

\begin{figure}
  \includegraphics[scale=0.4]{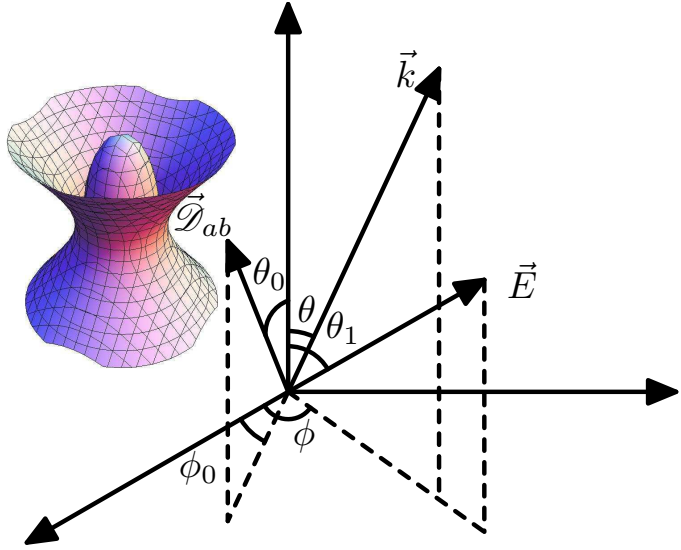}\\
\caption{The schematic diagram of the three vectors. inset:  The geometry of the
dispersion relations of the indefinite  medium}\label{scheme}
\end{figure}

Since the elements of permittivity(permeability) tensor of the metamaterials can be
negative or positive in different frequency range\cite{indefinite} we apply the term
\emph{indefinite }to anisotropic media in which not all of the principal components of
the ``$\tensor{\epsilon}$" tensor has same sign.  The geometry of the dispersion relation
in the indefinite media can be characterized by an ellipsoid or a hyperboloid as shown in
the inset to Fig.\ref{scheme}. The medium with hyperboloid geometry of the dispersion
relation is also called ``hyperbolic medium"(HM). The HM has been recently found to be
having many novel properties such as the superlens
effect\cite{hyperlen1,hyperlen2,hyperlen3}, slow-light effect\cite{hyper} and the  ``big
flash" of the photons in the HM by an optical metric signature phase
transition\cite{metric}. Moreover, some interesting  quantum optical properties(QOPs) of
the HM has also been found experimentally\cite{controlling} and
theoretically\cite{Broadband,Zhukovskiy,Ivan}, for instance, controlling spontaneous
emission with the HM\cite{controlling}, broadband Purcell effect\cite{Broadband}, the
dipole radiation and its enhancement near the surface of the HM\cite{Zhukovskiy,Ivan}.
There are also some studies on the QOPs of the indefinite  media   where the emphasis is
put on the left-handed property\cite{lefthanded} and the singularity of the density of
states\cite{Ultrahigh}.

The above theoretical analysis on the QOPs of the indefinite  media  are within framework
of the macroscopic electromagnetic wave theory and  the atom is approximated as a
dipole\cite{interference,Broadband,Zhukovskiy,Ivan} where the radiated power is exacted
from the Green's function of the system. However, the rate of spontaneous emission for a
two-level atom in the indefinite  media has not been investigated within the framework of
the quantum optics. In this letter we will explicitly  give  the expression of the decay
rate for a two-level atom in the indefinite media under the Weisskopf-Wigner
approximation. It indicates that for the ellipsoid case the intensity of the photons
coupled into different modes can be tuned by changing the direction of the matrix
element. For the hyperboloid case it is found that spontaneous emission in the HM can be
dramatically enhanced in comparison with the dielectric background, meanwhile, the
spontaneous emission exhibit the strong directivity and get the maximum in the asymptote
direction.

\section{model and formula}
To deal with the our problem  some theories of the  quantum optics of dielectric media is
needed. There are many schemes of  the electromagnetic quantization  for the different
types of media. One of them  is that the  quantities characterizing the dielectric such
as polarization field, magnetization field  or some quantities of their combination are
involved in the quantization procedure  and  the interaction between their  quanta and
the photon is taken into account\cite{Kheirandish,Amooshahi,Bruno}.  Another scheme is
that the permittivity and permeability  tensor  are  regarded as the  parameters  in the
field equation with  different gauges(gauge conditions) chosen  for the different
media\cite{Glauber,Lenac,bianisotropic}. In our case the medium is considered as
anisotropic, homogenous, non-dispersive and lossless  for simplicity. The quantization
scheme in Ref.\cite{bianisotropic} is used where the  medium is characterized by the
constitutive equations $\mathbf{D}(r)=\tensor{\epsilon}^{(1)}(r)\cdot \mathbf{E}(r)
+\tensor{\epsilon}^{(2)}(r)\cdot\mathbf{B}(r)$; $\mathbf{H}(r)=\tensor{\mu}^{(1)}(r)\cdot
\mathbf{E}(r) +\tensor{\mu}^{(2)}(r)\cdot\mathbf{B}(r)$. For our  case
$\tensor{\epsilon}^{(2)}(r)=\tensor{\mu}^{(1)}(r)=0;\; \tensor{\mu}^{(2)}(r)=1 $ and
\begin{equation}\label{eq:epsilon}
\tensor{\epsilon}^{(1)}(r)\equiv\tensor{\epsilon}=\epsilon_0
\left(\begin{array}{ccc}\epsilon_t &0&0\\0&\epsilon_t &0\\0 & 0
&\epsilon_L\\\end{array}\right)\end{equation}

where $\epsilon_L>0$  and  $\epsilon_t>0$ for the uniaxial anisotropic medium or
$\epsilon_L>0$  and  $\epsilon_t<0$ for the HM. According to the Maxwell's equations  the
dispersion relation for the medium characterized by Eq.(\ref{eq:epsilon})  in principal
axis coordinate system  is  expressed
as:\begin{equation}\label{eq:dispersion}\frac{k_z^2}{\epsilon_t}+\frac{k_x^2+k_y^2}{\epsilon_L}
=\left(\frac{\omega}{c}\right)^2 \end{equation} The geometry of the
Eq.(\ref{eq:dispersion}) in the K space represent a hyperboloid or an ellipsoid that
depends on the sign of $\epsilon_t$ where $\epsilon_L>0$ is assumed. Here we are
interested in the decay rate for a two-level atom in the indefinite media. As the
excitation of the quantum electromagnetic field  the  photon is emitted form  an atom and
subsequently coupled into the field modes permitted by the dielectric environment. In
dielectric system the photon propagates in the fashion of the classical field modes of
the system and inversely interact with the atom. In the spirit of Einstein's original
model the total energy of the electromagnetic field mode in the dielectric should be
$\hbar\omega_k$\cite{Garrison}, which can be used to determine the field amplitude
 in the factor $g_k$. According to Ref.\cite{bianisotropic} the eigenvector field
in the dielectric should satisfy the following gauge condition and the eigenequations:
\begin{eqnarray}
&&\nabla\cdot\left[\tensor{\epsilon}(r)\cdot \vec{F}_k(r)\right]=0 \\
&& \nabla\times\nabla\times \vec{F}_k(r)=\omega_k^2
\tensor{\epsilon}(r)\cdot\vec{F}_k(r)
\end{eqnarray}
The quantized electric field can be expressed as $\hat{\vec{E}}(r,t)=\sum_k
\vec{e}_k F_k(r) e^{-i\omega_kt} \hat{a}_k +H.c.$ and the total  Hamiltonian
of the photon and the atom under the rotating-wave approximation  is
\begin{eqnarray}\label{Hamis}
\nonumber
\hat{\mathscr{H}}&=&\sum_k\hbar\omega_k \hat{a}_k^{\dag}\hat{a}_k
+\frac{1}{2}\hbar\nu\hat{\sigma}_z
\\
&+&\hbar\sum_kg_k(\hat{\sigma}_{+}\hat{a}_k+\hat{\sigma}_{-}\hat{a}_k^{\dag})
\end{eqnarray}
where $\hat{\sigma}_z=|a\rangle\langle a|-|b\rangle\langle b|,
\;\hat{\sigma}_{+}=|a\rangle\langle b|,\;\hat{\sigma}_{-}=|b\rangle\langle a|$,
$|a\rangle,|b\rangle$
is the excited  and ground states of atom with  eigenvalues $E_a,E_b$  and
\begin{equation}\label{gk}
g_k=-\frac{\vec{\mathscr{D}}_{ab}\cdot \vec{e}_kF_k(0)}{\hbar}
\end{equation}
$ \vec{\mathscr{D}}_{ab} $ is the matrix element of  atom's dipole between
states $\vec{\mathscr{D}}_{ab} \equiv - \langle a|e\hat{\vec{R}}
|b\rangle=-\langle b|e\hat{\vec{R}} |a\rangle$.

For simplicity the Hamiltonian Eq.(\ref{Hamis}) can be expressed as  in the
interaction\cite{Scully} picture \begin{equation}\label{HamiI}
\mathscr{H}_I=\hbar\sum_k\left[g_k^*\hat{\sigma}_{+}\hat{a}_ke^{i(\omega_0-\omega_k)t}
+g_k\hat{\sigma}_{-}\hat{a}_k^{\dag}e^{-i(\omega_0-\omega_k)t} \right]
\end{equation} where $E_a-E_b=\hbar\omega_0$.
The state vector of the  composite system of the photon and atom  including
the vacuum state $|0\rangle$  is
$|\psi(t)\rangle=c_a|a,0\rangle+\sum_kc_{b,k}|b,1_k\rangle$.   With the
Weisskopf-Wigner approximation the decay rate of the atom in the anisotropic
medium can be expressed as the following integration
\begin{equation}\label{gamma} \Gamma= \frac{V}{(2\pi)^3}\int
\left.\left(|g_k(0)|^2 k^2 \frac{\partial \omega}{\partial
k}^{-1}\right)\right|_{\omega=\omega_0}\sin\theta d\theta d\phi\end{equation}
where the dispersion relation of the modes is written as
$\omega=f(k,\theta,\phi)$ and $k$ is $k=f^{-1}(\omega,\theta,\phi)$. Here the
eigen electric field is assumed as
$\vec{E}_k(r)=\vec{F}_k(r);\;\vec{H}_k=\frac{1}{i\mu_0\omega}\nabla\times\vec{E}_k$.
That a single photon is coupled into the eigen-mode  require the total energy
of the mode $
U=\frac{1}{2}\displaystyle\int\limits_{\mbox{v}}\left(\vec{E_k}\cdot\tensor{\epsilon}\cdot\vec{E}_k
+\mu_0|\vec{H}_k|^2\right) d^3x=\hbar\omega$, which determine the amplitude
$F_k(0)$ under the box normalization  condition.

\subsection{Spontaneous decay rate in the medium with dispersion geometry of
ellipsoid}

In order to check  the validity of the  formulism, the case $\epsilon_t>0$ is firstly
considered where the medium become uniaxial anisotropic with an ellipsoidal dispersive
geometry. In the medium there are usually two types of  eigen-modes  named as
extraordinary wave and ordinary wave which accommodate the emitted photons form the atom.
To this end we have to explore the amplitudes and the energy of the two modes in detail.
The eigen electric field is supposed as  the plane wave
$\vec{E}_k(r)=\vec{E}_{k0}e^{i\vec{k}\cdot\vec{r}}e^{i\omega(\vec{k})t}$ and
$\vec{H}_k(r)=\vec{H}_{k0}e^{i\vec{k}\cdot\vec{r}}e^{i\omega(\vec{k})t}$ where the
complex vector amplitudes are to be determined.  Substitute the expression into the
Maxwell Equations and after some algebra we get the two sets of  dispersion and
polarization relations

\begin{eqnarray}
                & &k = \frac{\omega}{c}  \sqrt{\epsilon_t}\label{Trans} \\
                & &\mbox{with } E_z=0 \mbox{ and } k_xE_{k0x}+k_yE_{k0y}=0
                \nonumber\\
               &
               &\frac{k_x^2+k_y^2}{\epsilon_L}+\frac{k_z^2}{\epsilon_t}=\left(\frac{\omega}{c}\right)^2\label{Logni}\\
             & &\mbox{with } H_z=0 \mbox{ and }
             \frac{E_{k0x}}{E_{k0y}}=\frac{k_x}{k_y}
             \nonumber
          \end{eqnarray}
In the spherical coordinate system   the dispersion relations can be uniformlly written
as: $\omega(k)=f(\theta) c k$, where $f(\theta)=\frac{1}{\sqrt{\epsilon_t}}$ for the
transversal mode Eq(\ref{Trans}), and
$f(\theta)=\sqrt{\frac{\sin^2\theta}{\epsilon_L}+\frac{\cos^2\theta}{\epsilon_t}}$ for
the lognitudinal mode Eq(\ref{Logni}).  For the transverse mode  the total energy in the
volume  V and the  relative amplitude are
 \begin{eqnarray}
  & & U_T=\epsilon_0(E_{k0}^{T})^2\epsilon_t V=\hbar\omega_0 \nonumber\\
   & &    E_{k0}^{T}=\sqrt{\frac{\hbar\omega_0}{V\epsilon_0\epsilon_t}}
 \end{eqnarray}

For the lognitudinal mode  the  corresponding quantities :

\begin{eqnarray}
  & & U_L=\frac{V \epsilon _0^2 \mu _0
\omega_0 ^2 (E_{k0}^L)^2 \epsilon _L^2 \epsilon _t^2}{k_z^2 \epsilon _L^2+\epsilon
_t^2
\left(k_x^2+k_y^2\right)}=\omega_0 \hbar \nonumber \\
   & &    E_{k0}^L=\left[\frac{\omega_0  \hbar  \left(\cos (2 \theta )
    \left(\epsilon _L^2-\epsilon _t^2\right)
    +\epsilon _L^2+\epsilon _t^2\right)}{2 f^2 V \epsilon_0 \epsilon _L^2 \epsilon
    _t^2}\right]^{1/2}
 \end{eqnarray}

In the principal axis spherical  coordinate system there are three vectors to be
identified. They are the vectorial  transition matrix element
$\vec{\mathscr{D}}_{ab}=(\mathscr{D}_{ab},\theta_0,\phi_0)$, the electric field vector
$\vec{E}_{k0}=(E_{k0},\theta_1,\phi_1)$ and the wave vector $\vec{k}=(k,\theta,\phi)$
which is shown in Fig. \ref{scheme}.   The factor in Eq.(\ref{gamma}) is explicitly given
\begin{equation}
    |g_k|^2=\frac{\mathscr{D}_{ab}^2 E_{k0}^2\cos^2 \theta_{0,1} }{\hbar^2}
\end{equation}
 where $\theta_{0,1}$ is the angle between $\vec{\mathscr{D}}_{ab}$ and
 $\vec{E}_{k0}$.
 According to the geometrically  relations of the vectors $\vec{\mathscr{D}}_{ab}$
 and  $\vec{E}_{k0}$  and the gauge condition
 $\vec{k}\cdot\tensor{\epsilon}\cdot\vec{E}_{k0}=0$ , we get the equations
 $\cos\theta_{0,1}  =\sin\theta_0 \sin\theta_1\cos(\phi_0-\phi_1)
 +\cos\theta_0\cos\theta_1$ and  $\epsilon_L\cos\theta _1\cos\theta
 +\epsilon_t\sin\theta_1 \sin\theta\cos\left(\phi_1-\phi\right)=0$.  In term of
 Eq.(\ref{Trans}),Eq.(\ref{Logni}),  It is noted that  $\phi_1=\phi+\frac{\pi}{2}$
 for the  transversal  mode and $\phi_1=\phi$ for the lognitudial  mode.  With these
 conditions  we  get the  factor $c_q=\cos^2\theta_{0,1}$ for the different modes
  \begin{eqnarray}& & c_q^T=\sin ^2\theta_0\sin ^2\left(\phi _0-\phi \right)
  \\& & c_q^L=\frac{\left[\epsilon_t\cos\theta_0-\epsilon_L \sin\theta_0\cot\theta
  \cos\left(\phi _0-\phi \right)\right]^2}{\epsilon_L^2\cot^2\theta+\epsilon_t^2}
  \end{eqnarray}

After the implementation of the Eq.(\ref{gamma}) we get the decay rate  for the two
modes
\begin{subequations}\label{tga}
\begin{eqnarray}
   & &
   \Gamma^{T}=\frac{\mathscr{D}_{ab}^2\omega_0^3\sqrt{\epsilon_t}\sin^2\theta_0}{4\pi
   c^3
   \epsilon_0\hbar }\label{gat} \\
  & & \Gamma^{L}=\frac{\mathscr{D}_{ab}^2
  \omega_0^3\left(-\cos\left(2\theta_0\right)
  \left(\epsilon_L-4 \epsilon_t\right)+\epsilon_L+4 \epsilon_t\right)}{24 \pi   c^3
  \epsilon_0\hbar\sqrt{\epsilon_t}}\label{gaL}\\
  &&\Gamma=\frac{\mathscr{D}_{ab}^2\omega_0^3\left(\cos\left(2\theta_0\right)\left(\epsilon_t-\epsilon
  _L\right)+\epsilon_L+7\epsilon_t\right)}{24\pi c^3
  \epsilon_0\hbar\sqrt{\epsilon_t}}\label{ga}
\end{eqnarray}
\end{subequations}

\begin{figure}[htb]
  \includegraphics[scale=1] {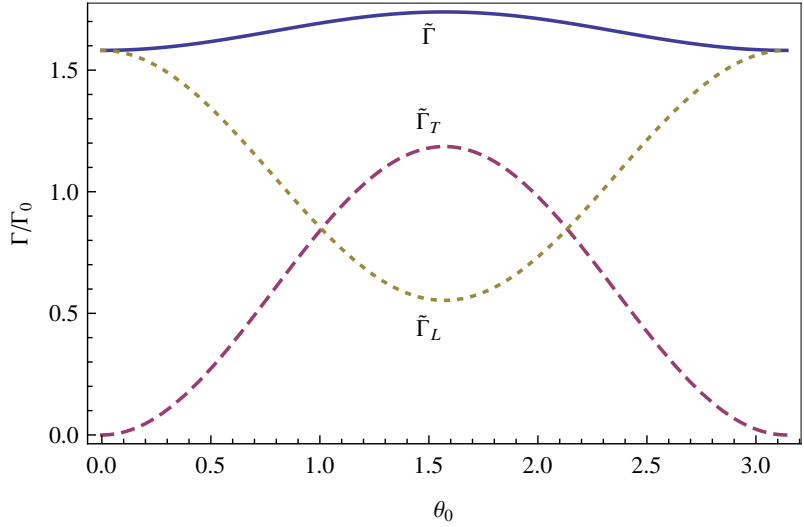}\\
\caption{The decay rate of atom  in  medium  with
$\epsilon_t=2.5,\epsilon_L=3.5$ relative to the case for the vacuum
$\tilde{\Gamma}=\Gamma/\Gamma_0$. Doted line: the decay rate for the transverse
mode. Dashed line: the decay rate for the  lognitudinal  mode.  Solid line: the total
decay
rate
}\label{gas}
\end{figure}

When $\epsilon_L=\epsilon_t>0$, Eq.(\ref{ga}) reduce as $\frac{\mathscr{D}_{ab}^2
\omega_0^3 \sqrt{\epsilon_t}}{3 \pi c^3 \epsilon_0 \hbar }=\Gamma(\epsilon_t)$, which
give the decay rate of the atom in the homogeneous isotropic medium with permittivity
$\epsilon_t$.   The decay rate relative to the vacuum $\Gamma/\Gamma_0=\tilde{\Gamma }$
in the case $\epsilon_t=2.5,\epsilon_L=3.5$ is shown in Fig.\ref{gas} for the different
modes.  Form the Fig.\ref{gas}  it is found that when the matrix element vector
$\vec{\mathscr{D}}_{ab}$ is parallel to the $k_z$ axis($\theta_0=0$), the atom can get
the maximal coupling with the lognitudial  mode of the system where $E_z\neq0$ and much
more photons emitted form the atom is coupled into the mode. Meanwhile, the transverse
mode get the no coupling with $\vec{\mathscr{D}}_{ab}$  for $E_z=0$  and
$\tilde{\Gamma}_T=0$.  In the case,
$\tilde{\Gamma}=\tilde{\Gamma}_L=1.58=\sqrt{\epsilon_t}$, which the anisotropic medium
behave as the isotropic medium with permittivity $\epsilon_t$ for the atom's spontaneous
emission.  With the increase of $\theta_0$, $\tilde{\Gamma}_T$ also increase due to the
enhancement of the coupling with the transverse mode  and $\tilde{\Gamma}_L$ decrease due
to the reduction of the coupling with the lognitudial  mode.  When
$\theta_0=\frac{\pi}{2}$ where $\vec{\mathscr{D}}_{ab}$  is perpendicular to the $k_z$
axis $\tilde{\Gamma}_T$ get the maximum  and $\tilde{\Gamma}_L$ get the minimum.
According to Eq.(\ref{ga})  $\Gamma$ get the maximum($\epsilon_t<\epsilon_L$) or the
minimum($\epsilon_t>\epsilon_L$): $\Gamma_m=\frac{\mathscr{D}_{ab}^2\omega_0^3\left(
2\epsilon_L+6\epsilon_t\right)}{24\pi c^3\epsilon_0\hbar\sqrt{\epsilon_t}}$ at
$\theta_0=\frac{\pi}{2}$.  These results can be used to control the intensity of the
different modes from the spontaneous emission by tuning $\theta_0$.

\subsection{Spontaneous decay rate in the medium with dispersion geometry of
hyperboloid }

For the case $\epsilon_t<0,\epsilon_L>0$, Eq.(\ref{eq:epsilon}) indicates the hyperboloid
geometry of the dispersion relation.  Since $\epsilon_t<0$, the branch
$k=\frac{\omega}{c}\sqrt{\epsilon_t}$ can only exist with evanescent wave. This mode can
not carry the energy away from the atom for the spontaneous emission  and the
contribution to the decay rate is ignored for simplicity.  For the branch
$\frac{k_x^2+k_y^2}{\epsilon_L}+\frac{k_z^2}{\epsilon_t}=\left(\frac{\omega}{c}\right)^2$
the decay rate $\Gamma$ is proportional to the following expression:
\begin{equation}\label{cosx}
 \Gamma= \beta\int_{-1}^{1} dx
    \frac{ x^2 \epsilon _L^2 \sin ^2 \theta_0 -2 \left(x^2-1\right)
    \epsilon_t^2\cos^2 \theta_0  }{\left(x^2-\epsilon_u\right){}^{5/2}}
\end{equation}

where $x=\cos\theta$, $\epsilon_u=\frac{\epsilon_t}{\epsilon_t-\epsilon_L}$,
$\beta=\frac{D^2 \omega^3 \left(\frac{\epsilon_L \epsilon_t}{\epsilon
_L-\epsilon_t}\right){}^{5/2}}{8\pi c^3 \epsilon_0 \hbar\epsilon_L^2 \epsilon _t^2}$

\begin{figure}
  \includegraphics[scale=1.5]{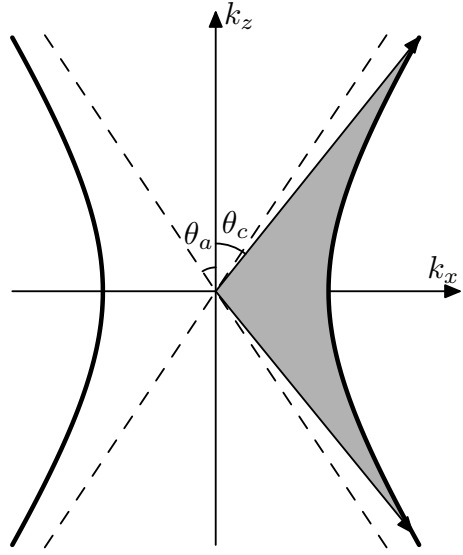}
\caption{The schematic diagram of  integration region  for the hyperbolic medium
where $\epsilon_t<0
$,$\epsilon_L>0$}\label{hy}
\end{figure}

It is noted that under the Weisskopf-Wigner approximation the integration
Eq.(\ref{gamma}) is actually  calculated on the equal-frequency surface. For the
ellipsoid Eq.(\ref{gamma}) $\theta$ integrates over the range $[0,\pi]$, while for the
hyperboloid $\theta$ integrates  over the range $[\theta_a,\pi-\theta_a]$, as  shown in
Fig.\ref{hy} where $\theta_a$ is the polar angle of the asymptote. In term of
Eq.(\ref{cosx}) there are two poles $\pm\sqrt{\epsilon_u}$ the positions of which on the
axis depend on the parameters $\epsilon_t,\epsilon_L$. When $\epsilon_t>\epsilon_L>0$,
$\epsilon_u>1$  and $\epsilon_L>\epsilon_t>0$, $\epsilon_u<0$ the poles
$\pm\sqrt{\epsilon_u}$ are out of the range $[-1,1]$ or on the imaginary axis. It enable
Eq.(\ref{cosx}) to be calculated and the result is given in Eq.(\ref{tga}). However, for
the hyperboloid case where $\epsilon_t<0,\epsilon_L>0$, $0<\epsilon_u<1$ the poles lie in
the range $[-1,1]$  the integration diverges.  This divergence is the manifestation of
the change of the topology from the ellipsoid to the  hyperboloid and it also indicate
that  hyperboloid of the dispersion relation is only the  perfect effective medium
approximation of some composite materials e.g. photonics crystal, metamaterials under the
long wave-length limit. Therefor, some cutoff methods have to be introduced for the
calculation of Eq.(\ref{cosx}) in the case $\epsilon_t<0,\epsilon_L>0$.  To this end it
is defined that $x_a=\cos\theta_a=\sqrt{\epsilon_u}$  and $\cos\theta_c=x_c\equiv\alpha
x_a$, where $\pm x_c$ is the new integration limits for Eq.(\ref{cosx}) with condition
$x_c<x_a$. The integration range of the variable $\theta$ for the cutoff is marked in the
shadowed region in Fig.\ref{hy}. When the coefficient $\alpha=1$,  the integration limit
approach to  the two polar poles $\pm x_a$. Under the cutoff approximation we get the
decay rate $\Gamma$ as follows:
\begin{widetext}
\begin{equation}\label{gamh}
 \Gamma_H=\Gamma_0\frac{\alpha ^3 \left[\epsilon_L^3-\epsilon_L\cos^2\theta_0
 \left(-4
 \epsilon _L \epsilon _t+\epsilon _L^2+6 \epsilon _t^2\right)\right]-6 \alpha
 \epsilon _L
 \epsilon _t \cos ^2\theta_0\left(\epsilon _L-\epsilon _t\right)}{4 \left(\alpha
 ^2-1\right)^{3/2} \left(\epsilon _L-\epsilon _t\right){}^2 \sqrt{\frac{\epsilon _L
 \epsilon
 _t}{\epsilon _L-\epsilon _t}}}
\end{equation}
where $\Gamma_0=\frac{\mathscr{D}_{ab}^2 \omega_0^3 }{3 \pi c^3 \epsilon_0
\hbar }$
\end{widetext}

\begin{figure}
  \includegraphics[scale=1] {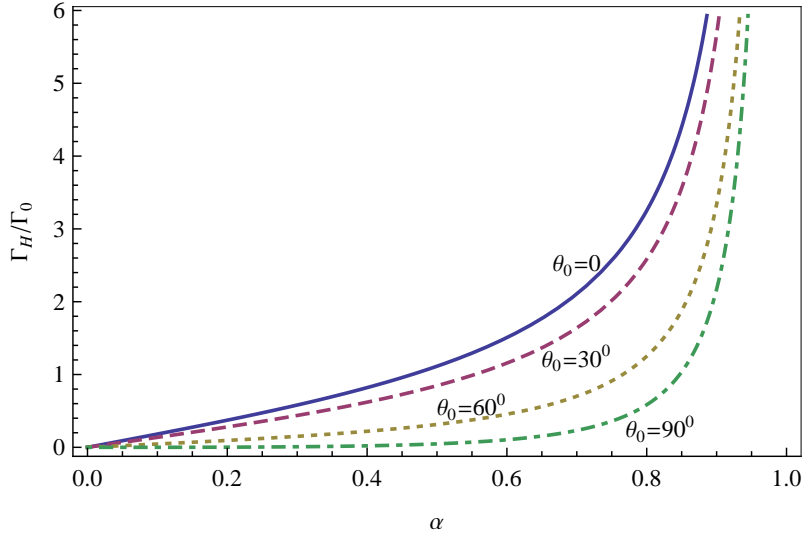}
\caption{The relative decay rate of atom  in  medium  with
$\epsilon_t=-2.5,\epsilon_L=3.5$ relative to the case for the vacuum
  according to parameter $\alpha$
for different angles $\theta_0$ }\label{gaa}
\end{figure}

In Fig.\ref{gaa} we give the $\tilde{\Gamma}_H=\Gamma/\Gamma_0$ vs the parameter $\alpha$
for the different $\theta_0$ with $\epsilon_t=-2.5,\epsilon_L=3.5$. It is obviously found
that $\tilde{\Gamma}_H$ increases on the whole when $\alpha$ approach $1$. It is noted
that when $\alpha$ close 1 the $\tilde{\Gamma}_H$ increases more sharply. It is because
that more large $\alpha$ is, more modes with  high $k$ vectors are involved in the
spontaneous emission. In HM the modes with high $k$ points usually have large density of
states  and  more photons emitted from the atom can be accommodated. Moreover, since the
photons is mainly coupled into the lognitudial mode, $\tilde{\Gamma}_H$  get more large
value when the matrix element vector $\vec{\mathscr{D}}_{ab}$ get the small angle
$\theta_0$ with $z$ axis which enhance the coupling factor $g_k$.

%
%
\begin{figure}
  \includegraphics[scale=1] {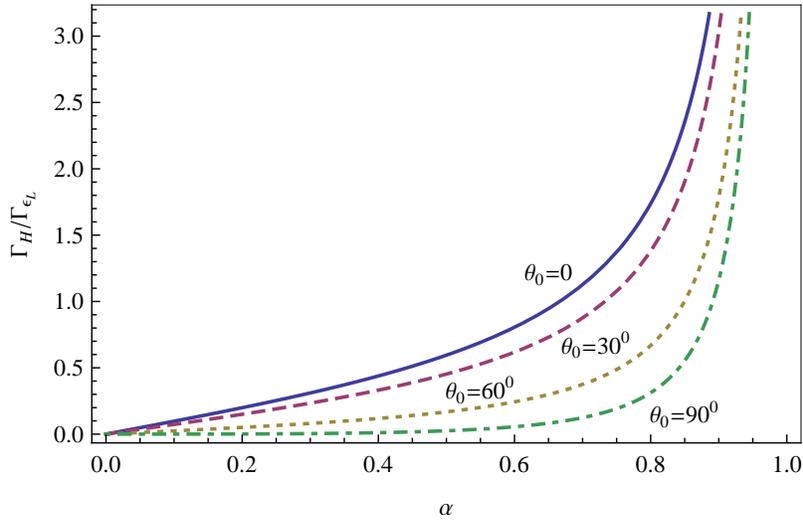}
\caption{The relative decay rate of atom  in  medium  with
$\epsilon_t=-2.5,\epsilon_L=3.5$ to the case for the background $\epsilon_L=3.5$
 according to parameter $\alpha$
for different angles $\theta_0$ }\label{gaaL}
\end{figure}

\begin{figure}
  \includegraphics[scale=1]{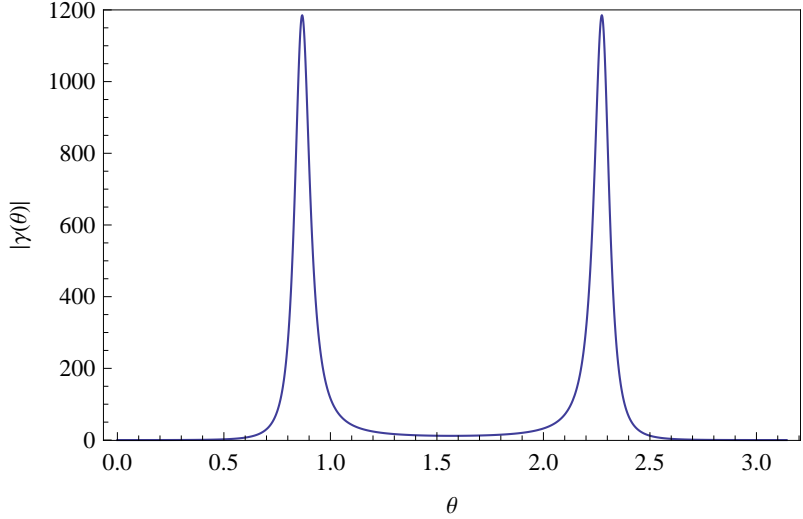}
\caption{The amplitude of the integrand $\gamma(\theta)$ corresponding to $\theta$
with
parameters
$\epsilon_t=-2.5+0.5 i,\epsilon_L=3.5$}\label{fgp}
\end{figure}

To explore the effect of the enhancement of the spontaneous emission we compare the decay
rate $\Gamma_H$ with the decay rate $\Gamma_{\epsilon_L}$, the decay rate in the
background medium with permittivity $\epsilon_L$. The relative decay rate
$\tilde{\Gamma}_H=\Gamma_H/\Gamma_{\epsilon_L}$ corresponding to parameter $\alpha$ is
given  in Fig.\ref{gaaL}.   It is easily found that $\tilde{\Gamma}_H$ can be lager than
one only if $\alpha$ is sufficiently large in spite of the difference of $\theta_0$,
which mean that the more like the strict hyperbolic medium the real materials behave, the
more the spontaneous emission is enhanced relative to the background medium. According to
Eq.(\ref{gamh}) there are $\lim\limits_{\alpha\rightarrow 1} \tilde{\Gamma}_H=\infty$,
which implicate that the perfect hyperbolic medium is the an idealization model of  some
real composite materials.  Besides the enhancement of the spontaneous emission the
directivity is also worthy of being noted. To this end the Eq.(\ref{cosx}) can be
rewritten as $\Gamma=\beta\int_0^{\pi} \gamma(\theta) d\theta$. Considering the
divergence, a small imaginary part is added to $\epsilon_t$. Fig.\ref{fgp} gives the
amplitude of the integrand $\gamma(\theta)$ with $\epsilon_t=-2.5+0.5 i,\epsilon_L=3.5$.
There are two obvious peaks in the positions corresponding to the directions of the
asymptotes. This strong directivity of the spontaneous emission has been noted
experimentally and used to design the single gun\cite{hyperlen2,Gunn}.

%
%

\section{discussion}

We explore the spontaneous emission of the two-level atom  in the homogeneous anisotropic
medium where the dispersion geometry exhibit as an ellipsoid or a hyperboloid and the
corresponding  decay rate $\Gamma$ in detail under the Weisskopf-Wigner approximation.
For the ellipsoid case, there are two kinds of modes that contribute the decay rate
$\Gamma$ and to some degree the medium with ellipsoid dispersion provides two type of
mode space to  accommodate the emitted photons. Moreover, the polar angle $\theta_0$ also
be used to change the intensity of  the photons coupled into the different modes.  When
$\epsilon_L=\epsilon_t$ the obtained formula Eq.(\ref{ga}) reduce to the case of the
isotropic medium, which justify the validity of our model.

\begin{figure}
\includegraphics[scale=1] {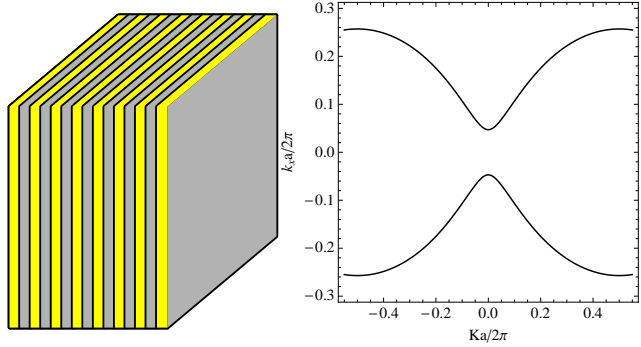}
\caption{Left:Schematic view  of the ``layered" metamaterials. Right: the
corresponding
equvi-frequency contour}\label{kg}
\end{figure}

When the above model is applied to the hyperboloid case  the divergence is encountered
and the cutoff for the integration variable $\theta$ (equivalently, the wave vector $K$)
is introduced for the approximation of the real materials.  Though the transverse
radiating wave mode degenerates into the evanescent wave mode, the enhanced spontaneous
emission relative to the background medium $\epsilon_L$ can be obtained only if  the
dispersion geometry of the  real materials is sufficiently  close to the strict
hyperboloid. It mean that more high $k$ modes which corresponds lager parameter $\alpha$
get involved in the coupling of the photons.  In  real world, the hyperbolic medium is
used as the perfect model on the some materials with periodic micro-structure  where band
structure exhibit the  approximative hyperbolic geometry in the small range of $k$, e.g.
the hyperbolic metamaterials. The simplest of them is the one dimensional photonic
crystal including  the metal layer as shown in Fig. \ref{kg} together with the
equifrequency contour. Generally, in the long wavelength limit  the system can be
approximated as the hyperbolic medium within the effective medium theory. It is expected
that when the more small the dimension of the lattice $a$ is, the more close to the
perfect hyperbolic medium the system is and  more enhancement of the spontaneous emission
can be achieved.

The physical mechanism of the enhancement on the spontaneous emission in HM can be
understood from the perspective of density of states.  According to the Fermi's golden
rule the one-to-many transition probability per unit of time depends not only  on the
matrix element but  the  density of final states $\rho(\omega)$ as well. When  the
geometry of physical dispersion relation of the medium changes  form the ellipsoid to the
hyperboloid,  due to the change of the topological property the density of states
diverges in the lossless continuous hyperbolic medium limit: $\rho(\omega)\approx
\frac{K^3_{cut }}{12\pi^2}\left|
\frac{\epsilon_L}{\epsilon_t}\left(\frac{1}{\epsilon_t}\frac{d\epsilon_t}
{d\omega}-\frac{1}{\epsilon_L}\frac{d\epsilon_L}{d\omega}\right)\right|$ where $K_{cut}$
is the momentum cutoff\cite{metric}. $K_{cut}$ is defined by either metamaterial
structure scale or by losses. It is the occurrence of the large $\rho$ that enable the
transition probability to be increased dramatically.  Even the loss and the dispersion of
the $\epsilon(\omega)$  in the materials is taken into account the spontaneous emission
is expected to be largely enhanced.


\begin{acknowledgments}
This work is supported by the NSFC (Grant No. 11004212, 11174309, 60877067 and 60938004,
the STCSM (Grant No. 11ZR1443800).
\end{acknowledgments}

\bibliography{texbib}

\clearpage

\end{document}